\documentclass[12pt]{article}
\usepackage{amssymb,amsfonts}
\usepackage{epsf,epsfig}
\begin{document}
\baselineskip 18pt

\title{Second cluster integral from the spectrum of an infinite $XXZ$ spin chain}
\author{P.~N.~Bibikov\thanks{bibikov@PB7855.spb.edu}}
\date{\it V.~A.~Fock Institute of Physics Saint-Petersburg State
University, Russia}
%\affiliation{Saint-Petersburg State University}
\maketitle

\vskip5mm

\begin{abstract}
First and second terms of the low-temperature cluster expansion for the free energy density of a magnetically polarized $XXZ$ spin chain is obtained
within the propagator approach suggested by E. W. Montroll and J. C. Ward. All the calculations employ only one- and two-magnon infinite-chain spectrums.
In the $XXX$-point the result reproduces the well known S. Katsura's formula obtained 50 years ago by finite-chain calculations.
\end{abstract}

\maketitle

\section{Introduction}

It is well known that the low-temperature thermodynamics of a {\it gapped} spin chain is governed by its low-lying spectrum \cite{1,2,3}.
The most studied example is the $XXZ$ ferromagnetic spin chain whose low-lying excitations are magnons, kinks and anti-kinks \cite{1,4}).
Among the other 1D gapped spin systems are rung-dimerized (rung-singlet) spin ladder \cite{3,5}, $S=1$ ferromagnetic chain
in magnetic field \cite{6}, alternating-spin ferromagnetic chain in magnetic field \cite{7}, easy-axis ferromagnetic zigzag chain in magnetic
field \cite{8} and so on. For all of these models the corresponding physical Hilbert space is an infinite direct sum of eigenspaces
\begin{equation}
{\cal H}^{(phys)}=\oplus_{Q=0}^{\infty}{\cal H}^{(Q)},
\end{equation}
where the subspace ${\cal H}^{(0)}$ is one-dimensional and generated by the ground state $|\emptyset\rangle$ while the sectors ${\cal H}^{(Q)}$
with $Q>0$ are the $Q$-magnon sectors. Expansion (1) is equivalent to an existence of a magnon number operator $\hat Q$
which satisfies the following conditions
\begin{eqnarray}
&&[\hat H,\hat Q]=0,\\
&&\hat Q\Big|_{{\cal H}^{(Q)}}=Q,
\end{eqnarray}
where $\hat H$ is the Hamiltonian.

In the special case when additionally to (2) and (3) the ground state of the system is an infinite tensor product of local ground states associated
with the sites of the chain
\begin{equation}
|\emptyset\rangle=\dots|\emptyset\rangle_{n-1}\otimes|\emptyset\rangle_n\otimes|\emptyset\rangle_{n+1}\dots\equiv\otimes_n|\emptyset\rangle_n,
\end{equation}
the corresponding one- and two-magnon states may be obtained by standard analysis \cite{9,10,11,12}. The $Q$-magnon states with $Q>2$ are
known only for a rare class of {\it integrable} models \cite{13,14}. However if the temperature and energy gaps $E_{gap}^{(Q)}$
related to the sectors ${\cal H}^{(Q)}$ satisfy the system of inequalities
\begin{equation}
k_BT,E_{gap}^{(1)}\ll E_{gap}^{(2)}\ll E_{gap}^{(Q)},\quad Q>2,
\end{equation}
then the thermodynamics is governed by the one-magnon spectrum being slightly corrected by the two-magnon one.
Since these corrections depend on the magnon-magnon interaction their account may be very useful.

Cluster expansion \cite{14,15,16,17,18,19} is a rigorous and consistent approach to the systems whose spectrum satisfy (1) and (5).
For a spin chain with $N$ sites it is based on the representation of the {\it partition function} as a sum over all $Q$-sectors \cite{14,15}
\begin{equation}
Z(T,N)=1+\sum_{Q=1}^{Q_{max}(N)}Z_Q(T,N),\qquad Z_Q(T,N)=\sum_{|\nu\rangle\in{\cal B}^{(Q)}}{\rm e}^{-\beta E_{\nu}},
\end{equation}
where ${\cal B}^{(Q)}$ is an eigenbasis of ${\cal H}^{(Q)}$. The parameter $Q_{max}(N)$ is the maximal number of particles which
may be excited in the chain with $N$-sites ($Q_{max}(\infty)=\infty$). In the thermodynamical limit $N\rightarrow\infty$ Eq. (6) results in
the following representation
\begin{equation}
f(T)=\sum_{Q=1}^{\infty}f_Q(T),
\end{equation}
for the free energy density
\begin{equation}
f(T)\equiv-k_BT\lim_{N\rightarrow\infty}\frac{\log{Z(T,N)}}{N}.
\end{equation}
The first two terms of the expansion (7) are
\begin{eqnarray}
f_1(T)&=&-k_BT\lim_{N\rightarrow\infty}\frac{Z_1(T,N)}{N},\\
f_2(T)&=&-k_BT\lim_{N\rightarrow\infty}\frac{2Z_2(T,N)-Z_1^2(T,N)}{2N}.
\end{eqnarray}

Formally the cluster expansion for spin chains is similar to the analogous one in the theory of imperfect
gases \cite{16,17,18,19} where the formulas (6) and (7) are replaced by
\begin{equation}
Z(T,V,\mu)=1+\sum_{Q=1}^{\infty}Z_Q(T,V)z^Q,
\end{equation}
and
\begin{equation}
f(T,\mu)\equiv-k_BT\lim_{V\rightarrow\infty}\frac{\log{Z(T,V,\mu)}}{V}=\sum_{Q=1}^{\infty}f_Q(T)z^Q.
\end{equation}
Here $V$ is a gas volume, $\mu$ is a chemical potential, $Z(T,V,\mu)$ is the {\it grand partition function}, $f(T,\mu)$ is the
corresponding free energy density, and
\begin{equation}
z\equiv{\rm e}^{\beta\mu},
\end{equation}
is an activity. However the convergence conditions for both the expansions are quite different.
Eqs. (11) and (12) give the expansion in powers of activity which converges well only at
\begin{equation}
z\ll1\Longleftrightarrow\beta\mu\ll0.
\end{equation}
In its turn the chemical potential is related to the gas concentration $n$ according to the well known formula \cite{19}
\begin{equation}
n(T,\mu)=-\frac{\partial f(T,\mu)}{\partial\mu}.
\end{equation}
In a spin chain all finite temperature excitations (magnons, their bound states e .t. c.) are {\it thermally excited} and hence all their chemical
potentials are equal to zero \cite{19} (really Eqs. (6), (7) may be obtained from (11), (12) by taking $z=1$). In this case the criteria (14) become
irrelevant and should be replaced by the system (5), because each term $Z_Q(T,N)$ of the sum (6) is proportional to the factor
${\rm e}^{-\beta E_{gap}^{(Q)}}$. That is why the expansion (6), (7) is a {\it low-temperature} expansion, while the expansion (11), (12) due to the
well known theorem \cite{19}
\begin{equation}
\lim_{T\rightarrow\infty}\mu=-\infty,
\end{equation}
better works at {\it high temperatures}.

A derivation of $f_1(T)$ needs only a knowledge of the magnon dispersion being a rather simple procedure. Postulating a phenomenological quadratic form
of the magnon dispersion M. Troyer, H. Tsunetsugu, D. ${\rm W\ddot urtz}$ in their seminal paper \cite{3} obtained $f_1(T)$ and the low-temperature
asymptotics for various thermodynamical quantities (magnetic susceptibility, heat capacity and so on) of a general 1D gapped system.
For the majority of systems $E_{gap}^{(1)}\ll E_{gap}^{(2)}$ and the low-temperature thermodynamics is governed by $f_1(T)$, while the term $f_2(T)$
is responsible for small corrections. However there is a number of compounds \cite{20,21,22} for which $E_{gap}^{(2)}\ll E_{gap}^{(1)}$ and
the term $f_2(T)$ becomes really actual. Fortunately its derivation needs only a knowledge of the two-particle spectrum which in the 1D case
and under conditions (1)-(4) may be readily obtained even for a wide class of nonintegrable systems \cite{23}.

To the author knowledge until now a derivation of $f_2(T)$ was performed only for the $XXX$ or Heisenberg chain long ago by S. Katsura \cite{15}.
Later this result was reproduced by M. Takahashi \cite{14} within a powerful approach suitable however only to integrable models.
Within the Katsura's method the limits (9) and (10) are calculated directly under a detailed study of one- and two-magnon
finite-$N$ spectrums. But although the finite-$N$ one-magnon problem is rather simple even the two-magnon case is just cumbersome \cite{24}.
At the same time since the free energy density $f(T)$ is a characteristic of an infinite chain it is natural to suppose that it may be obtained
directly from the infinite-$N$ spectrum which is quite more simple than the finite-$N$ one \cite{25,26}.

In the present paper developing the alternative {\it propagator} approach to cluster expansion \cite{18} and employing only infinite-$N$
wave functions we obtain $f_1(T)$ and $f_2(T)$ avoiding any finite-$N$ calculations and reproducing the Katsura's result in the isotropic $XXX$ point.

\section{One- and two-magnon states on the infinite chain}

We take Hamiltonian of the infinite chain $XXZ$ ferromagnet in the following form
\begin{equation}
\hat H=-\sum_n\Big[\frac{J}{2}\Big({\bf S}^+_n{\bf S}^-_{n+1}+{\bf S}^-_n{\bf S}^+_{n+1}\Big)+
J_z\Big({\bf S}^z_n{\bf S}^z_{n+1}-\frac{1}{4}\Big)+\frac{\gamma h}{2}\Big({\bf S}_n^z+{\bf S}_{n+1}^z-1\Big)\Big].
\end{equation}
Here ${\bf S}_n^z$ and ${\bf S}_n^{\pm}={\bf S}_n^x\pm i{\bf S}_n^y$ are the usual spin-1/2 operators.
and
\begin{equation}
J,J_z,\gamma\geq0,\qquad h>0.
\end{equation}

The corresponding spin-polarized ground state has the form (4) with
\begin{equation}
|\emptyset\rangle_n=|\uparrow\rangle_n,
\end{equation}
where by $|\uparrow\rangle_n$ and $|\downarrow\rangle_n$ are denoted spin polarized local states corresponding to $n$-th site.
General one- and two-magnon states have the forms
\begin{equation}
|1\rangle=\sum_n\psi_n^{(1)}{\bf S}^-_n|\emptyset\rangle,\qquad
|2\rangle=\sum_{n_1<n_2}\psi_{n_1,n_2}^{(2)}{\bf S}^-_{n_1}{\bf S}^-_{n_2}|\emptyset\rangle,
\end{equation}
where the wave functions satisfy the ${\rm Schr\ddot odinger}$ equations
\begin{eqnarray}
&&(J_z+\gamma h)\psi_n^{(1)}-\frac{J}{2}\Big(\psi_{n-1}^{(1)}+\psi_{n+1}^{(1)}\Big)=E\psi_n^{(1)},\\
&&2(J_z+\gamma h)\psi_{n_1,n_2}^{(2)}-\frac{J}{2}\Big(\psi_{n_1-1,n_2}^{(2)}+\psi_{n_1+1,n_2}^{(2)}+\psi_{n_1,n_2-1}^{(2)}+\psi_{n_1,n_2+1}^{(2)}\Big)
=E\psi_{n_1,n_2}^{(2)},\nonumber\\
&&n_2-n_1>1,\nonumber\\
&&(J_z+2\gamma h)\psi_{n,n+1}^{(2)}-\frac{J}{2}\Big(\psi_{n-1,n+1}^{(2)}+\psi_{n,n+2}^{(2)}\Big)=E\psi_{n,n+1}^{(2)},
\end{eqnarray}
or in short notations
\begin{equation}
H^{(1)}\psi^{(1)}=E\psi^{(1)},\qquad H^{(2)}\psi^{(2)}=E\psi^{(2)}.
\end{equation}

Eqs. (23) have the following complete systems of solutions related to one-magnon and scattering or bound two-magnon states
\begin{eqnarray}
&&\psi_n^{(1)}(k)={\rm e}^{ikn},\\
&&\psi_{n_1,n_2}^{(2,scatt)}(k,\kappa)={\rm e}^{ik(n_1+n_2)/2}\varphi_{n_2-n_1}^{(scatt)}(k,\kappa),
\quad\psi_{n_1,n_2}^{(2,bound)}(k)={\rm e}^{ik(n_1+n_2)/2}\varphi_{n_2-n_1}^{(bound)}(k),\quad
\end{eqnarray}
where
\begin{equation}
\varphi_{n}^{(scatt)}(k,\kappa)=\frac{A(k,\kappa){\rm e}^{i\kappa n}-A(k,-\kappa){\rm e}^{-i\kappa n}}{\sqrt{A(k,\kappa)A(k,-\kappa)}},\quad
\varphi_{n}^{(bound)}(k)=\frac{\sqrt{J_z^2-J^2\cos^2{k/2}}}{J\cos{k/2}}\Big(\frac{J}{J_z}\cos{\frac{k}{2}}\Big)^n,
\end{equation}
and
\begin{equation}
A(k,\kappa)\equiv J\cos{\frac{k}{2}}-J_z{\rm e}^{-i\kappa},\qquad \kappa\in{\mathbb R}.
\end{equation}
According to the relations
\begin{equation}
\varphi_n^{(scatt)}(k,-\kappa)=-\varphi_n^{(scatt)}(k,\kappa),\quad \varphi_n^{(scatt)}(k,0)=\varphi_n^{(scatt)}(k,\pi)\equiv0,
\end{equation}
one may put
\begin{equation}
0<\kappa<\pi.
\end{equation}
The bound states exist only under the condition
\begin{equation}
\Big|\frac{J}{J_z}\cos{\frac{k}{2}}\Big|<1.
\end{equation}

The corresponding dispersions are
\begin{eqnarray}
&&E_{magn}(k,h)=J_z-J\cos{k}+\gamma h,\nonumber\\
&&E_{scatt}(k,\kappa,h)=E_{magn}(k/2-\kappa,h)+E_{magn}(k/2+\kappa,h),\nonumber\\
&&E_{bound}(k,h)=J_z-\frac{J^2}{J_z}\cos^2{\frac{k}{2}}+2\gamma h.
\end{eqnarray}

For evaluation of the cluster expansion we shall need system of relations
\begin{eqnarray}
&&\sum_n\bar\psi_n^{(1)}(k)\psi_n^{(1)}(\tilde k)=2\pi\delta(k-\tilde k),\\
&&\int_0^{2\pi}dk\bar\psi_n^{(1)}(k)\psi_{\tilde n}^{(1)}(k)=2\pi\delta_{n \tilde n},\\
&&\sum_{n_1<n_2}\bar\psi_{n_1,n_2}^{(2,scatt)}(k,\kappa)\psi_{n_1,n_2}^{(2,scatt)}(\tilde k,\tilde\kappa)
=(2\pi)^2\delta(k-\tilde k)\delta(\kappa-\tilde\kappa),\\
&&\sum_{n_1<n_2}\bar\psi_{n_1,n_2}^{(2,scatt)}(k,\kappa)\psi_{n_1,n_2}^{(2,bound)}(\tilde k)=0,\\
&&\sum_{n_1<n_2}\bar\psi_{n_1,n_2}^{(2,bound)}(k)\psi_{n_1,n_2}^{(2,bound)}(\tilde k)=2\pi\delta(k-\tilde k),\\
&&\frac{1}{(2\pi)^2}\int_0^{2\pi}dk\int_0^{\pi}d\kappa\bar\psi^{(2,scatt)}_{n_1,n_2}(k,\kappa)\psi^{(2,scatt)}_{\tilde n_1,\tilde n_2}(k,\kappa)\nonumber\\
&&+\frac{1}{2\pi}\int_0^{2\pi}dk\Theta\Big(J_z^2-J^2\cos^2{\frac{k}{2}}\Big)\bar\psi^{(2,bound)}_{n_1,n_2}(k)
\psi^{(2,bound)}_{\tilde n_1,\tilde n_2}(k)=\delta_{n_1,\tilde n_1}\delta_{n_2,\tilde n_2}.
\end{eqnarray}
Here $\Theta(x)=0$ for $x\leq0$ and $\Theta(x)=1$ for $x>0$.

Eqs. (32) and (33) directly follow from (24). Using representations (25) Eqs. (34)-(37) may be represented in equivalent forms
\begin{eqnarray}
&&\sum_{n=1}^{\infty}\bar\varphi_n^{(scatt)}(k,\tilde\kappa)\varphi_n^{(scatt)}(k,\kappa)=2\pi\delta(\kappa-\tilde \kappa),\quad\\
&&\sum_{n=1}^{\infty}\bar\varphi_n^{(scatt)}(k,\kappa)\varphi_n^{(bound)}(k)=0,\\
&&\sum_{n=1}^{\infty}\bar\varphi_n^{(bound)}(k)\varphi_n^{(bound)}(k)=1,\\
&&\frac{1}{2\pi}\int_0^{\pi}\bar\varphi_n^{(scatt)}(k,\kappa)\varphi_{\tilde n}^{(scatt)}(k,\kappa)d\kappa=\delta_{n\tilde n}
-\Theta\Big(J_z^2-J^2\cos^2{\frac{k}{2}}\Big)\nonumber\\
&&\cdot\Big[1-\Big(\frac{J}{J_z}\cos{\frac{k}{2}}\Big)^2\Big]
\Big(\frac{J}{J_z}\cos{\frac{k}{2}}\Big)^{(n+\tilde n-2)}.
\end{eqnarray}

Eqs. (39) and (40) may be readily proved directly. Eq. (38) follows from the formula
\begin{eqnarray}
&&\bar\varphi_n^{(scatt)}(k,\kappa)\varphi_n^{(scatt)}(k,\tilde\kappa)=\frac{2}{\sqrt{A(k,\kappa)A(k,-\kappa)
A(k,\tilde\kappa)A(k,-\tilde\kappa)}}\nonumber\\
&&\cdot\Big[J^2\cos^2{\frac{k}{2}}\Big(\cos{(\kappa-\tilde\kappa)n}-\cos{(\kappa+\tilde\kappa)n}\Big)
-2JJ_z\cos{\frac{k}{2}}\cos{\frac{\kappa+\tilde\kappa}{2}}\Big(\cos{(\kappa-\tilde\kappa)(n-1/2)}\nonumber\\
&&-\cos{(\kappa+\tilde\kappa)(n-1/2)}\Big)
+J_z^2\Big(\cos{(\kappa-\tilde\kappa)(n-1)}-\cos{(\kappa+\tilde\kappa)(n-1)}\Big)\Big].
\end{eqnarray}
Finitely with the use of the representation
\begin{equation}
\bar\varphi_n^{(scatt)}(k,\kappa)\varphi_{\tilde n}^{(scatt)}(k,\kappa)=W_{n,\tilde n}(k,\kappa)+W_{n,\tilde n}(k,-\kappa),
\end{equation}
where
\begin{equation}
W_{n,\tilde n}(k,\kappa)={\rm e}^{i\kappa(n-\tilde n)}-\frac{J_z{\rm e}^{-i\kappa}-J\cos{k/2}}{J_z{\rm e}^{i\kappa}-
J\cos{k/2}}{\rm e}^{i\kappa(n+\tilde n)},
\end{equation}
Eq. (41) reduces to the form
\begin{equation}
\frac{1}{2\pi}\int_{-\pi}^{\pi}d\kappa W_{n,\tilde n}(k,\kappa)=\delta_{n\tilde n}
-\Theta\Big(J_z^2-J^2\cos^2{\frac{k}{2}}\Big)\Big(1-\frac{J^2\cos^2{k/2}}{J_z^2}\Big)\Big(\frac{J\cos{k/2}}{J_z}\Big)^{n+\tilde n-2},
\end{equation}
which may be readily proved by direct integration.

\section{Propagator approach to the cluster expansion}

Let $H^{(1)}_N$ and $H^{(2)}_N$ be finite-$N$ analogues of $H^{(1)}$ and $H^{(2)}$. Corresponding one- and two-magnon eigenspaces
${\cal H}_N^{(1)}$ and ${\cal H}_N^{(2)}$ are generated by $N$- and $N(N-1)/2$-dimensional vectors of the form
\begin{equation}
|1\rangle=\sum_{n=1}^N\psi_n^{(1)}{\bf S}^-_n|\emptyset\rangle,\qquad
|2\rangle=\sum_{1\leq n_1<n_2\leq N}\psi_{n_1,n_2}^{(2)}{\bf S}^-_{n_1}{\bf S}^-_{n_2}|\emptyset\rangle,
\end{equation}
where in this case $|\emptyset\rangle=\otimes_{n=1}^N|\uparrow\rangle_n$.

Let $\psi_n^{(1)}(\mu_1)$ and $\psi_{n_1,n_2}^{(2)}(\mu_2)$ be the sets (enumerated by the indexes $\mu_1$ and $\mu_2$) of normalized wave functions
related to some eigenbases in ${\cal H}_N^{(1)}$ and ${\cal H}_N^{(2)}$. Orthonormality and completeness of the eigenbases result in the following
system of relations
\begin{eqnarray}
\sum_{n=1}^N\bar\psi^{(1)}_n(\mu_1)\psi^{(1)}_n(\tilde\mu_1)=\delta_{\mu_1\tilde\mu_1},&&
\sum_{1\leq n_1<n_2\leq N}\bar\psi^{(2)}_{n_1,n_2}(\mu_2)\psi^{(2)}_{n_1,n_2}(\tilde\mu_2)=\delta_{\mu_2\tilde\mu_2},\\
\sum_{\mu_1}\bar\psi^{(1)}_n(\mu_1)\psi^{(1)}_{\tilde n}(\mu_1)=\delta_{n \tilde n}&&
\sum_{\mu_2}\bar\psi^{(2)}_{n_1,n_2}(\mu_2)\psi^{(2)}_{\tilde n_1,\tilde n_2}(\mu_2)=\delta_{n_1 \tilde n_1}\delta_{n_2 \tilde n_2}.
\end{eqnarray}
According to (23) and (48) the $N\times N$ and $N(N-1)/2\times N(N-1)/2$ matrices
\begin{eqnarray}
&&K^{(1)}_{n,\tilde n}(\beta,h,N)=\sum_{\mu_1}{\rm e}^{-\beta E_{\mu_1}}\bar\psi_n^{(1)}(\mu_1)\psi_{\tilde n}^{(1)}(\mu_1),\\
&&K^{(2)}_{n_1n_2,\tilde n_1\tilde n_2}(\beta,h,N)=
\sum_{\mu_2}{\rm e}^{-\beta E_{\mu_2}}\bar\psi_{n_1,n_2}^{(2)}(\mu_2)\psi_{\tilde n_1,\tilde n_2}^{(2)}(\mu_2),
\end{eqnarray}
satisfy at $\beta>0$ differential equations
\begin{equation}
\frac{\partial K^{(1)}(\beta,h,N)}{\partial\beta}+H^{(1)}_NK^{(1)}(\beta,h,N)=0,\quad
\frac{\partial K^{(2)}(\beta,h,N)}{\partial\beta}+H^{(2)}_NK^{(2)}(\beta,h,N)=0,
\end{equation}
and $\beta\rightarrow0$ conditions
\begin{equation}
K^{(1)}_{n,\tilde n}(0,h,N)=\delta_{n\tilde n},\quad K^{(2)}_{n_1n_2,\tilde n_1\tilde n_2}(0,h,N)=\delta_{n_1\tilde n_1}\delta_{n_2\tilde n_2}.
\end{equation}
Moreover, as it follows from (47)
\begin{equation}
Z_1(T,h,N)=\sum_{n=1}^NK^{(1)}_{n,n}(\beta,h,N),\quad Z_2(T,h,N)=\sum_{1\leq n_1<n_2\leq N}K^{(2)}_{n_1n_2,n_1n_2}(\beta,h,N),
\end{equation}
and hence according to (9), (10) and translation invariance of the $N=\infty$ model
\begin{eqnarray}
&&f_1(T,h)=-k_BT\lim_{N\rightarrow\infty}\frac{1}{N}\sum_{n=1}^NK^{(1)}_{n,n}(\beta,h,N)=-k_BT\lim_{N\rightarrow\infty}K^{(1)}_{0,0}(\beta,h,N),\\
&&f_2(T,h)=-k_BT\lim_{N\rightarrow\infty}\Big\{\frac{1}{N}\sum_{1\leq n_1<n_2\leq N}\Big[K^{(2)}_{n_1n_2,n_1n_2}(\beta,h,N)-
\Big(K^{(1)}_{0,0}(\beta,h,N)\Big)^2\Big]\nonumber\\
&&-\frac{1}{2}\Big(K^{(1)}_{0,0}(\beta,h,N)\Big)^2\Big\}.
\end{eqnarray}

In the $N\rightarrow\infty$ limit the infinite dimensional operators
\begin{equation}
K^{(1)}(\beta,h)=\lim_{N\rightarrow\infty}K^{(1)}(\beta,h,N),\quad K^{(2)}(\beta,h)=\lim_{N\rightarrow\infty}K^{(2)}(\beta,h,N),
\end{equation}
should satisfy analogous to (51) and (52) differential equations
\begin{equation}
\frac{\partial K^{(1)}(\beta,h)}{\partial\beta}+H^{(1)}K^{(1)}(\beta,h)=0,\quad
\frac{\partial K^{(2)}(\beta,h)}{\partial\beta}+H^{(2)}K^{(2)}(\beta,h)=0,
\end{equation}
and $\beta\rightarrow0$ conditions
\begin{equation}
K^{(1)}_{n,\tilde n}(0,h)=\delta_{n\tilde n},\quad K^{(2)}_{n_1n_2,\tilde n_1\tilde n_2}(0,h)=\delta_{n_1\tilde n_1}\delta_{n_2\tilde n_2}.
\end{equation}
Due to the translation invariance of the infinite chain Eqs. (54) and (55) may be reduced to
\begin{eqnarray}
&&f_1(T,h)=-k_BTK^{(1)}_{0,0}(\beta,h),\\
&&f_2(T,h)=-k_BT\Big\{\sum_{n=1}^{\infty}\Big[K^{(2)}_{0n,0n}(\beta,h)-\Big(K^{(1)}_{0,0}(\beta,h)\Big)^2\Big]
-\frac{1}{2}\Big(K^{(1)}_{0,0}(\beta,h)\Big)^2\Big\}.
\end{eqnarray}

Fortunately the operators $K^{(1)}(\beta,h)$ and $K^{(2)}(\beta,h)$ may be obtained
directly with the use of one- and two-magnon infinite chain wave functions. Namely
\begin{equation}
K^{(1)}_{n,\tilde n}(\beta,h)=\frac{1}{2\pi}\int_0^{2\pi}dk{\rm e}^{-\beta E_{magn}(k,h)}\bar\psi^{(1)}_n(k)\psi^{(1)}_{\tilde n}(k),
\end{equation}
and
\begin{equation}
K^{(2)}(\beta,h)=K^{(2,scatt)}(\beta,h)+K^{(2,bound)}(\beta,h),
\end{equation}
where
\begin{eqnarray}
K^{(scatt)}_{n_1n_2,\tilde n_1\tilde n_2}(\beta,h)=\frac{1}{(2\pi)^2}\int_0^{2\pi}dk\int_0^{\pi}d\kappa
{\rm e}^{-\beta E_{scatt}(k,\kappa,h)}
\bar\psi^{(scatt)}_{n_1,n_2}(k,\kappa)\psi^{(scatt)}_{\tilde n_1,\tilde n_2}(k,\kappa),\nonumber\\
K^{(bound)}_{n_1n_2,\tilde n_1\tilde n_2}(\beta,h)=\frac{1}{2\pi}\int_0^{2\pi}dk\Theta(J_z^2-J^2\cos^2{k/2}){\rm e}^{-\beta E_{bound}(k,h)}
\bar\psi^{(bound)}_{n_1,n_2}(k)\psi^{(bound)}_{\tilde n_1,\tilde n_2}(k).
\end{eqnarray}
or according to (25) and (31)
\begin{eqnarray}
&&K^{(scatt)}_{n_1n_2,\tilde n_1\tilde n_2}(\beta,h)=\frac{1}{(2\pi)^2}\int_0^{2\pi}dk{\rm e}^{ik(\tilde n_1+\tilde n_2-n_1-n_2)/2}\int_0^{\pi}d\kappa{\rm e}^{-\beta [E_{magn}(k/2-\kappa)+E_{magn}(k/2+\kappa)]}\nonumber\\
&&\cdot\bar\varphi^{(scatt)}_{n_2-n_1}(k,\kappa)\varphi^{(scatt)}_{\tilde n_2-\tilde n_1}(k,\kappa),\\
&&K^{(bound)}_{n_1n_2,\tilde n_1\tilde n_2}(\beta,h)=\frac{1}{2\pi}\int_0^{2\pi}dk{\rm e}^{ik(\tilde n_1+\tilde n_2-n_1-n_2)/2}\Theta(J_z^2-J^2\cos^2{k/2}){\rm e}^{-\beta E_{bound}(k,h)}\nonumber\\
&&\bar\varphi^{(bound)}_{n_2-n_1}(k)\varphi^{(bound)}_{\tilde n_2-\tilde n_1}(k).
\end{eqnarray}
Eqs. (57) and (58) now directly follow from (23) and (48)

A substitution of (24) into (61) gives
\begin{equation}
K^{(1)}_{n,\tilde n}(\beta,h)=\frac{1}{2\pi}\int_0^{2\pi}dk{\rm e}^{ik(\tilde n-n)-\beta E_{magn}(k,h)}.
\end{equation}
Hence according to (59) and (66)
\begin{equation}
f_1(T,h)=-\frac{k_BT}{2\pi}\int_0^{2\pi}dk{\rm e}^{-\beta E_{magn}(k,h)}.
\end{equation}
Evaluation of $f_2(T,h)$ is given in the next section.

\section{Evaluation of the second cluster integral}

Following Eqs. (60) and (62) we put
\begin{equation}
f_2(T,h)=f^{(scatt)}_2(T,h)+f^{(bound)}_2(T,h),
\end{equation}
where
\begin{eqnarray}
f^{(scatt)}_2(T,h)&=&-k_BT\Big\{\sum_{n=1}^{\infty}\Big[K^{(2,scatt)}_{0n,0n}(\beta,h)-
\Big(K^{(1)}_{0,0}(\beta,h)\Big)^2\Big]-\frac{1}{2}\Big(K^{(1)}_{0,0}(\beta,h)\Big)^2\Big\},\quad\,\,\\
f^{(bound)}_2(T,h)&=&-k_BT\sum_{n=1}^{\infty}K^{(2,bound)}_{0n,0n}(\beta,h).
\end{eqnarray}
According to (66)
\begin{equation}
\Big(K^{(1)}_{00}(T,h)\Big)^2=\frac{1}{(2\pi)^2}\int_0^{2\pi}dk_1\int_0^{2\pi}dk_2{\rm e}^{-\beta [E_{magn}(k_1,h)+E_{magn}(k_2,h)]},
\end{equation}
or passing to the new variables $k=k_1+k_2$ and $\kappa=(k_2-k_1)/2$
\begin{equation}
\Big(K^{(1)}_{00}(T,h)\Big)^2=\frac{1}{(2\pi)^2}\int_0^{2\pi}dk\int_{-\pi}^{\pi}d\kappa{\rm e}^{-\beta [E_{magn}(k/2-\kappa,h)+E_{magn}(k/2+\kappa,h)]}.
\end{equation}

Substituting (64) and (72) into (69) and taking into account that according to (26) and (27)
$\varphi_n^{(scatt)}(k,-\kappa)=-\varphi_n^{(scatt)}(k,\kappa)$ one gets
\begin{eqnarray}
&&f^{(scatt)}_2(T,h)=
\frac{k_BT}{8\pi^2}\int_0^{2\pi}dk\int_{-\pi}^{\pi}d\kappa{\rm e}^{-\beta[E_{magn}(k/2-\kappa,h)+E_{magn}(k/2+\kappa,h)]}\nonumber\\
&&\cdot\Big[\sum_{n=1}^{\infty}\Big(2-\bar\varphi^{(scatt)}_n(k,\kappa)\varphi^{(scatt)}_n(k,\kappa)\Big)+1\Big].
\end{eqnarray}
Following Eqs. (42) and (27)
\begin{eqnarray}
&&\sum_{n=1}^{\infty}\Big(2-\bar\varphi^{(scatt)}_n(k,\kappa)\varphi^{(scatt)}_n(k,\kappa)\Big)=
\frac{2}{A(k,\kappa)A(k,-\kappa)}\sum_{n=1}^{\infty}
\Big(J^2\cos^2{\frac{k}{2}}\cos{2\kappa n}\nonumber\\
&&-2JJ_z\cos{\frac{k}{2}}\cos{\kappa}\cos{\kappa(2n-1)}
+J_z^2\cos{2\kappa(n-1)}\Big)\nonumber\\
&&=\pi\Big(\delta(\kappa)+\delta(\kappa-\pi)\Big)
+J_z\Big(\frac{1}{J_z-J{\rm e}^{i\kappa}\cos{k/2}}+\frac{1}{J_z-J{\rm e}^{-i\kappa}\cos{k/2}}\Big)-1.
\end{eqnarray}
Hence
\begin{eqnarray}
&&f^{(scatt)}_2(T,h)=
\frac{k_BT}{4\pi}\int_0^{2\pi}dk{\rm e}^{-2\beta E_{magn}(k,h)}\nonumber\\
&&+\frac{J_zk_BT}{(2\pi)^2}\int_0^{2\pi}dk\int_{-\pi}^{\pi}\frac{{\rm e}^{-\beta E_{scatt}(k,\kappa,h)}(J_z-J\cos{k/2}\cos{\kappa})d\kappa}
{J_z^2-2JJ_z\cos{k/2}\cos{\kappa}+J^2\cos^2{k/2}},
\end{eqnarray}
or equivalently
\begin{eqnarray}
&&f^{(scatt)}_2(T,h)=
\frac{k_BT}{4\pi}\int_0^{2\pi}dk{\rm e}^{-2\beta E_{magn}(k,h)}\nonumber\\
&&+\frac{J_zk_BT}{(2\pi)^2}\int_0^{2\pi}dk\int_{-\pi}^{\pi}d\kappa\frac{{\rm e}^{-\beta[E_{magn}(k/2-\kappa,h)+E_{magn}(k/2+\kappa,h)]}}
{J_z-J{\rm e}^{-i\kappa}\cos{k/2}}.
\end{eqnarray}

Correspondingly according to Eqs. (65), (70) and (40)
\begin{equation}
f^{(bound)}_2(T,h)=-\frac{k_BT}{2\pi}\int_0^{2\pi}dk\Theta(J_z^2-J\cos^2{k/2}){\rm e}^{-\beta E_{bound}(k,h)}.
\end{equation}

At $J=J_z$ Eqs. (68), (75) and (77) reproduce the Katsura's result \cite{15}. At $J_z=0$ Eqs. (67) and (75) give the first two terms of expansion for
the free energy of $XX$-chain \cite{27}
\begin{equation}
f^{(XX)}(T,h)=-\frac{k_BT}{2\pi}\int_0^{2\pi}dk\log{\Big(1+{\rm e}^{-\beta E_{magn}(k,h)}\Big)},
\end{equation}
in powers of ${\rm e}^{-\beta E_{gap}}$.

At $J_z/J>1$ and $h=0$ the model has two fully polarized ground states. The numbers of one- and two-magnon states also redouble and an expression
similar to (6) result in a similar cluster expansion for free energy density in powers of ${\rm e}^{-\beta E_{gap}}$. It is naturally to suppose
however that this expansion is correct only for $J/J_z>0.6$, because at $J/J_z<0.6$ the low-temperature thermodynamic is governed by kink
and anti-kink states \cite{1},\cite{4}. In this context it is instructive to study the case $J=0$ when the Hamiltonian (17) turns into the
classical Izing Hamiltonian
\begin{equation}
\hat H_{Izing}=\frac{1}{4}\sum_nJ_z(1-\sigma_n\sigma_{n+1})+\gamma h(2-\sigma_n-\sigma_{n+1}),\quad\sigma_n=\pm1.
\end{equation}
At $h>0$ the low-lying excitations corresponding to (79) are non dispersive magnons (single reversed spins) with constant energy
\begin{equation}
E_{magn}=J_z+\gamma h.
\end{equation}
However at $h=0$ the lowest excitations are kinks and anti-kinks
\begin{eqnarray}
&&|kink,n\rangle=\Big(\otimes_{j=-\infty}^{n}|\uparrow\rangle\Big)\otimes\Big(\otimes_{j=n+1}^{\infty}|\downarrow\rangle\Big),\nonumber\\
&&|anti-kink,n\rangle=\Big(\otimes_{j=-\infty}^{n}|\downarrow\rangle\Big)\otimes\Big(\otimes_{j=n+1}^{\infty}|\uparrow\rangle\Big),
\end{eqnarray}
with energies
\begin{equation}
E_{kink}=\frac{J_z}{2}.
\end{equation}

Using the standard approach \cite{28} one may readily obtain the corresponding expression for the free energy
\begin{equation}
f^{(Izing)}(T,h)=-k_BT\log{\Big[\frac{1}{2}
\Big(1+{\rm e}^{-\beta\gamma h}+\sqrt{\Big(1-{\rm e}^{-\beta\gamma h}\Big)^2+4{\rm e}^{-\beta(J_z+\gamma h)}}\Big)\Big]}.
\end{equation}
In this case Eqs. (67), (68), (76) and (77) result in
\begin{equation}
f_1^{(Izing)}(T,h)=-k_BT{\rm e}^{-\beta E_{magn}},\quad
f_2^{(Izing)}(T,h)=k_BT{\rm e}^{-\beta E_{magn}}\Big(\frac{3}{2}{\rm e}^{-\beta E_{magn}}-{\rm e}^{-\beta\gamma h}\Big).
\end{equation}
As it follows from Eq. (84) at $h=0$ the suggested cluster expansion becomes inefficient because bound states have the same energy as a one-magnon
state and the condition
$|f_2^{(Izing)}(T,h)|\ll|f_1^{(Izing)}(T,h)|$ fails. Moreover in this case the free energy takes the form
\begin{equation}
f^{(Izing)}(T,0)=-k_BT\log{\Big[1+{\rm e}^{-\beta E_{kink}}\Big]},
\end{equation}
and should be expanded in powers of ${\rm e}^{-\beta E_{kink}}$.

Using the representation (76) we may obtain a compact representation for $f_2(T,h)$.
Taking in (76) $w\equiv{\rm e}^{i\kappa}$ one readily gets
\begin{equation}
\frac{J_z}{2\pi}\int_{-\pi}^{\pi}d\kappa\frac{{\rm e}^{-\beta[E_{magn}(k/2-\kappa,h)+E_{magn}(k/2+\kappa,h)]}}{J_z-J{\rm e}^{-i\kappa}\cos{k/2}}=
\frac{1}{2\pi i}\oint_{|w|=1}\frac{{\rm e}^{-\beta[2J_z+2\gamma h-J(w+1/w)\cos{k/2}]}}{w-{\rm e}^{i\kappa_b}}dw,
\end{equation}
where
\begin{equation}
{\rm e}^{i\kappa_b}\equiv\frac{J}{J_z}\cos{\frac{k}{2}}.
\end{equation}
Taking into account that according to (31) and (87)
\begin{equation}
E_{bound}(k,h)=E_{magn}(k/2-\kappa_b,h)+E_{magn}(k/2+\kappa_b,h),
\end{equation}
one readily reduces Eq. (86) to the form
\begin{eqnarray}
&&\frac{J_z}{2\pi}\int_{-\pi}^{\pi}d\kappa\frac{{\rm e}^{-\beta[E_{magn}(k/2-\kappa,h)+E_{magn}(k/2+\kappa,h)]}}{J_z-J{\rm e}^{-i\kappa}\cos{k/2}}=
\Theta(J_z^2-J\cos^2{k/2}){\rm e}^{-\beta E_{bound}(k)}\nonumber\\
&&+\lim_{\epsilon\to0}\frac{1}{2\pi i}\oint_{|w|=\epsilon}\frac{{\rm e}^{-\beta[2J_z+2\gamma h-J(w+1/w)\cos{k/2}]}}{J_zw-J\cos{k/2}}dw.
\end{eqnarray}
Eqs. (68), (76), (89) and (77) result in the following representation
\begin{equation}
f_2(T,h)=\frac{k_BT}{2\pi}\int_0^{2\pi}dk\Big[\frac{{\rm e}^{-2\beta E_{magn}(k,h)}}{2}+
\frac{J_z}{2\pi i}\lim_{\epsilon\to0}\oint_{|w|=\epsilon}\frac{{\rm e}^{-\beta[2J_z+2\gamma h-J(w+1/w)\cos{k/2}]}}{J_zw-J\cos{k/2}}dw\Big].
\end{equation}

\section{Summary and discussions}

In the present paper using the propagator approach of Montroll and Ward \cite{18} we obtained the first two terms of free energy density
cluster expansion  for 1D $XXZ$-ferromagnet (Eqs. (67), (68), (75) and (77)). All the calculations employed the infinite-chain spectrum only.
The suggested approach may be applied to other spin chains with known one- and two-magnon infinite-chain spectrums.
The result is also suitable for an antiferromagnetic $XXZ$ spin chain ($J<0$ in Eq. (17)) in a saturation magnetic field.
In the three special points (isotropic Heisenberg chain, $XX$-chain, and Izing chain in magnetic field) the obtained formulas coincide with the
well known results.
At zero magnetic field we suggest that the result should be physically adequate only in the Heisenberg-Izing region ($0.6<J/J_z<1$) where
the low-temperature thermodynamics is governed by the magnon spectrum.

\end{document}